\documentstyle[11pt,newpasp,twoside,psfig,wrapfig]{article}
\index{Arecibo}
\index{central compact objects}
\index{Chandra}
\index{isolated neutron stars!RX~J0720.4--3125}
\index{isolated neutron stars!RX~J1308.6+2127}
\index{isolated neutron stars!RX~J1605.3+3249}
\index{isolated neutron stars!RX~J1836.2+5925}
\index{isolated neutron stars!RX~J1856.5--3754}
\index{gamma-rays}
\index{GBT}
\index{Parkes}
\index{pulsar wind nebulae}
\index{pulsars}
\index{pulsars!braking indices}
\index{supernova remnants!associations with pulsars}
\index{pulsars!radio}
\index{pulsars!surveys}
\index{pulsars!J0205+6449}
\index{pulsars!J1124--5916}
\index{pulsars!J1747--2958}
\index{pulsars!J2229+6114}
\index{pulsar wind nebulae!3C58}
\index{pulsar wind nebulae!G54.1+0.3}
\index{pulsar wind nebulae!G292.0+1.8}
\index{pulsar wind nebulae!G359.23--0.82}
\index{pulsar wind nebulae!Kes~75}
\index{pulsar wind nebulae!G292.2--0.5}
\index{ROSAT}
\index{unidentified gamma-ray sources!associations with pulsars}

\def\edcomment#1{\iffalse\marginpar{\raggedright\sl#1\/}\else\relax\fi}
\reversemarginpar

\begin{document}
\title{Searches for Young Pulsars}
\author{Fernando Camilo}
\affil{Columbia Astrophysics Lab, Columbia University, 550 West 120th
Street, New York, NY 10027, USA}

\begin{abstract}
I review the results of radio and X-ray searches for pulsations from
young neutron stars, emphasizing work accomplished in the last five years.
I cover undirected searches, as well as directed searches of pulsar wind
nebulae, EGRET $\gamma$-ray sources, and also the search for pulsations
from ``isolated neutron stars'' (INSs) and ``central compact
objects''.
\end{abstract}

\section{Introduction}

With one supernova occurring in the Galaxy every $\sim100$ years,
it follows that young neutron stars are rare.  Because of the manifold
applications of their study, however, considerable effort continues to be
devoted to detecting young neutron stars.  In the age of \textit{Chandra}
and \textit{XMM}, the existence of a young magnetospherically active
neutron star can often be taken for granted even in the absence of
the detection of pulsations, most spectacularly via the imaging and
spectroscopy of the beautiful structures that form when the relativistic
pulsar winds are confined by the ambient pressure --- a point source
surrounded by such a pulsar wind nebula (PWN) is proof enough.  Still,
proceeding to extract maximum understanding from the multi-wavelength
observations of such systems requires knowledge of the neutron star
spin period $P$, its derivative, and derived quantities: the spin-down
luminosity $\dot E = 4\pi^2 I \dot P/P^3$, which ultimately is the
energy source for all that we detect from most pulsars/PWNe, where
the neutron star moment of inertia is $I \equiv 10^{45}$\,g\,cm$^2$;
the characteristic age $\tau_c = P/2\dot P$, usually considered to be an
upper limit on the real age, although because of unknown braking indices
of rotation $n$ (where $\dot \nu \propto - \nu^n$ and $\nu = 1/P$), the
real age can easily be up to about twice this value; and an estimate
of the surface magnetic dipole field strength, $B = 3.2\times10^{19}
(P\dot P)^{1/2}$\,G.  For this reason, this review will emphasize
searches for \textit{pulsations} from young neutron stars, summarizing
efforts made in this area in the past five years.

\section{How to Search for Pulsars}

In principle, pulsations can be searched throughout the entire
electromagnetic spectrum, from radio to $\gamma$-ray wavelengths.
In practice, for reasons having to do both with the spectral energy
distribution of pulsed emission and the available instrumentation, only
radio and X-ray have been successful bands for such searches (see, e.g.,
Chandler et al.\ 2001, for a search for $\gamma$-ray pulsations).

The original discovery of pulsars relied on the detection of individual
radio pulses with no need to account for their dispersive propagation.
Since the early days few pulsars have been discovered via detection
of single pulses (see Nice 1999 for an exception).  Nevertheless it is
important to keep in mind the case of the Crab pulsar, discovered via its
``giant pulses''.  The important point here is that giant (dispersed)
pulses can be exceptionally bright, detectable at up to $\sim$
Mpc distances!  See McLaughlin \& Cordes (2003), Johnston \& Romani
(these proceedings) and Cordes (these proceedings) for a discussion of
giant pulses.

The standard modern method to search for pulsars relies on the use of Fast
Fourier Transform-based techniques to analyze time series of up to several
million pulses.  Also, for long observations, ``acceleration search''
techniques can be be employed (e.g., Ransom 2001) to partially correct
for a changing pulse period caused by binary motion or a large $\dot P$
from a young pulsar.  The above applies to both radio and X-ray searches.
For radio searches, in addition, dispersive propagation must be corrected
for by ``de-dispersing'' the time samples from many narrow frequency
channels at a large ($\sim 100$--1000) number of trial dispersion measures
before searching the one-dimensional time series for periodic signals.

\section{Where to Search for Pulsars}

Search efforts consist of undirected/``all-sky'' surveys and directed
searches of promising objects.  The most successful undirected surveys
for young pulsars cover the Galactic plane (see Fig.~1 left), of which
the Parkes multibeam survey represents the state-of-the-art (see \S4).
However, even high-latitude searches can uncover nearby or high-velocity
young pulsars (e.g., McLaughlin et al.\ 2002).

Directed searches include those of the following targets:

\begin{enumerate}

\item PWNe: by definition these contain young pulsars; whether the pulsars
are beaming toward the Earth or are luminous enough for pulsations to
be detected is a question to be addressed by sensitive searches (see \S5.2).

\item Supernova remnants (SNRs): obviously a good place to look for young
pulsars --- but some are rather large, and not all contain neutron stars
(see, e.g., Kaplan et al., these proceedings).

\item $\gamma$-ray (EGRET) sources: a subset of these are expected to
be powered by young pulsars.  However the positional uncertainty of the
sources is usually very large, making blind searches of the error
boxes a difficult task (see, e.g., Nice \& Sayer 1997; Roberts et al.,
these proceedings).  The process is simplified if there is a good X-ray
candidate within the error box that can be followed with maximal effort
to search for pulsations (see \S5.1).

\item ``Unusual'' (X-ray) sources: 

\begin{itemize}

\item[-] \textit{ROSAT} soft sources (see \S6; Kaplan 2004).

\item[-] Central Compact Objects (see \S7; Pavlov et al.,
these proceedings).

\item[-] AXPs and SGRs (see Kaspi, these proceedings).  None have been
detected at radio wavelengths (e.g., Crawford et al.\ 2002).

\end{itemize}

\item Radio steep spectrum/polarized sources: the young pulsar B1951+32
in SNR CTB~80 was found in this manner (Strom 1987; Kulkarni et al.\ 1988).
This method has not produced results of late (e.g., Crawford et al.\ 2000;
Kaplan et al.\ 2000; Kouwenhoven 2000).

\end{enumerate}

\section{The Parkes Multibeam Survey of the Galactic Plane}

\begin{figure}
\centerline{
\psfig{file=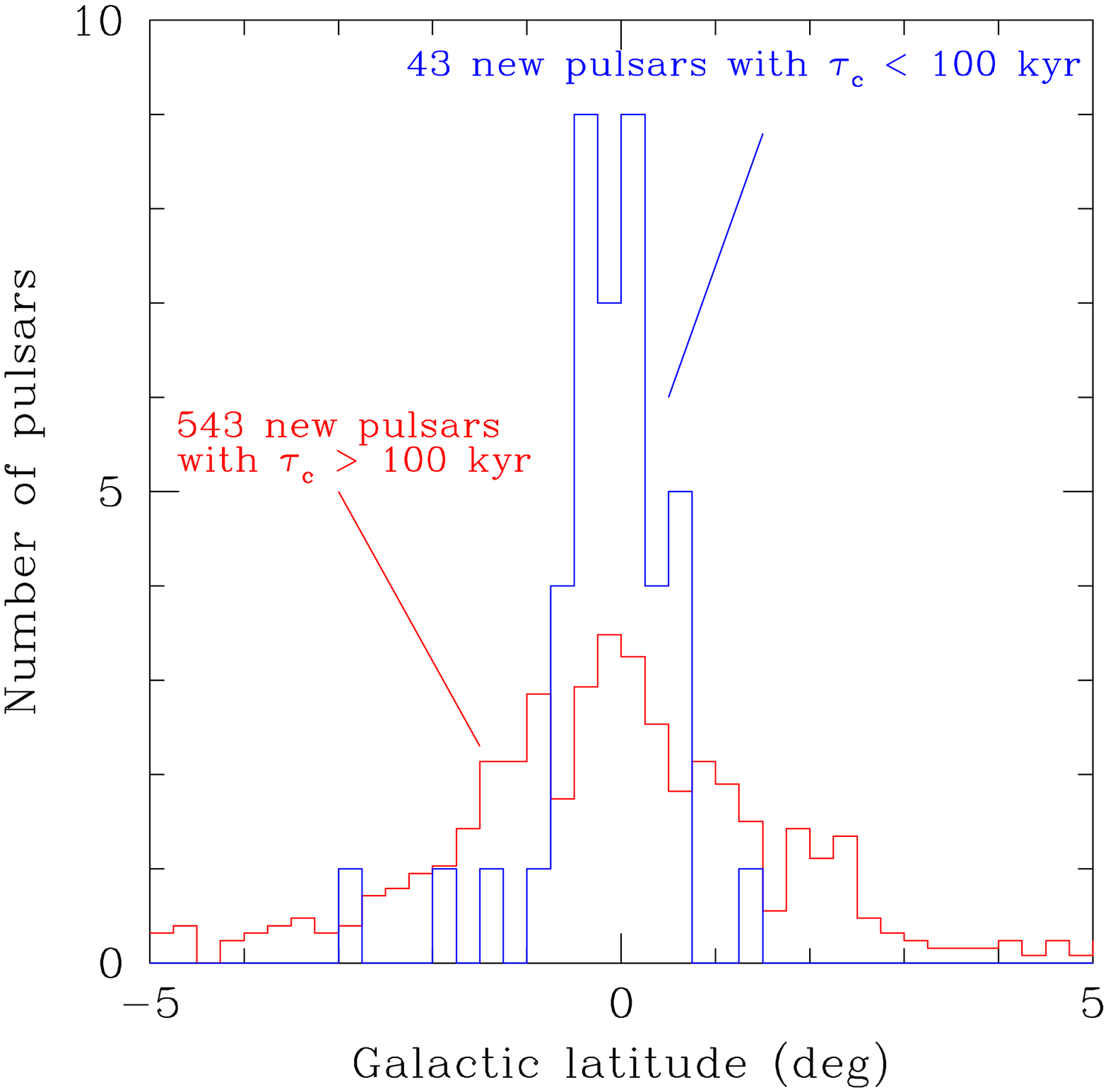,width=0.45\textwidth}
\psfig{file=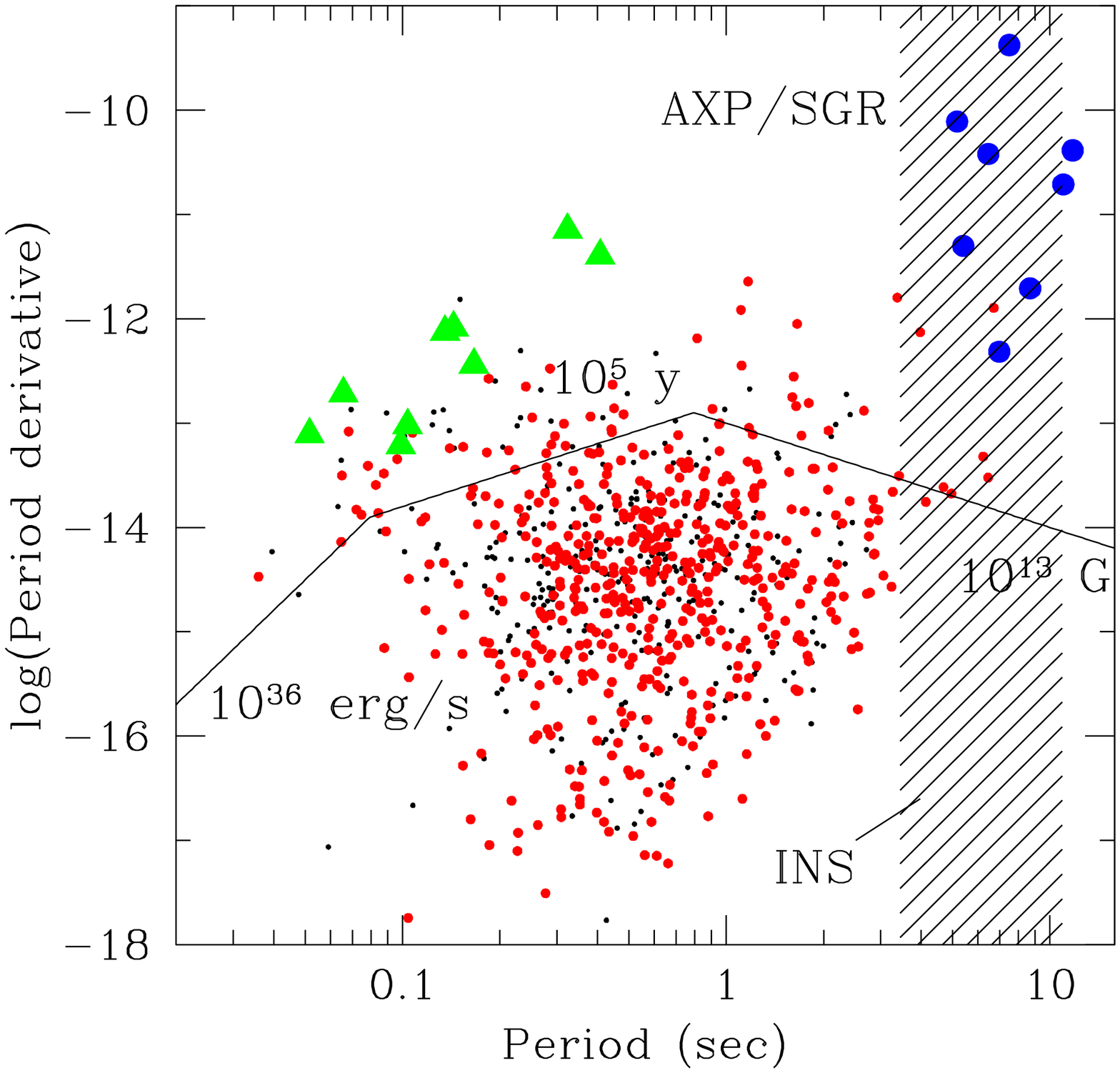,width=0.45\textwidth}}
\caption{\textit{Left:} Distribution in latitude for the pulsars
discovered in the Parkes multibeam survey that have a known characteristic
age: young pulsars have a much smaller scale-height than older ones
(both histograms have equal areas). \textit{Right:} The $P$--$\dot P$
diagram with three lines, above which lie the energetic, young, or high-$B$
pulsars; the period range of the INSs is indicated, and AXPs/SGRs are
shown at top right. Multibeam pulsars are represented by medium-size dots.
Nine young pulsars mentioned in \S4.1 and Table~\ref{tab:young} are
indicated at top left. }
\end{figure}

This search along the inner Galactic plane ($|b|<5\deg$; $260\deg < l <
50\deg$) has discovered $\sim 725$ new pulsars in an area where 330 were
previously known (e.g., Kramer et al.\ 2003).  Here we emphasize aspects
related to young pulsars.  The survey had a sensitivity comparable to
those of the directed SNR searches of the 1990s (Gorham et al.\ 1996;
Kaspi et al.\ 1996; Lorimer et al.\ 1998).

The survey has found 43 pulsars with $\tau_c < 10^5$\,yr, compared to
24 previously known.  Of these, 14 are ``Vela-like'' in that they are
energetic, with $\dot E > 10^{36}$\,ergs\,s$^{-1}$, doubling the number of
such pulsars known --- the kind that primarily power X-ray-bright PWNe.
On average the new pulsars are more distant than those previously known;
some are being observed with \textit{Chandra} and \textit{XMM} and it
remains to be seen what they may add to PWN phenomenology.

An unexpected haul from the survey has been the number of pulsars
with $B>10^{13}$\,G, 25 compared to nine previously known.  Some have
spin parameters indistinguishable from those of some AXPs (e.g.,
McLaughlin et al.\ 2003, and these proceedings).  However, their
emission characteristics are apparently entirely radio-pulsar-like,
with no particularly plentiful X-rays.

\subsection{Youngest multibeam pulsars; a serendipitous X-ray discovery}

\subsubsection{J1119--6127:}

Unusual in being very young ($\tau_c = 1600$\,yr) while having
a large $P=0.4$\,s.  The latter is a consequence of its large
$B=4\times10^{13}$\,G.  It is also one of only six pulsars with a known
braking index ($n=2.9$; Camilo et al.\ 2000).  The newly identified
(Crawford et al.\ 2001) X-ray-bright (Pivovaroff et al.\ 2001) shell SNR
G292.2--0.5 was subsequently found to be associated with the pulsar,
and a faint X-ray PWN has also been identified (Gonzalez \& Safi-Harb 2003).

\subsubsection{J1357--6429:}

Less exotic than J1119--6127, but still one of the 10 youngest
Galactic pulsars known, with $P=166$\,ms and $\tau_c=7300$\,yr.  It is
positionally coincident with the SNR candidate G309.8--2.6 (Camilo et al.,
in preparation).  Further radio and X-ray work are required in order to
characterize this system.

\subsubsection{J1846--0258:}

Discovered serendipitously by \textit{RXTE} within the composite
SNR Kes~75, with parameters comparable to those of J1119--6127:
$P=0.3$\,s, $\tau_c=720$\,yr (Gotthelf et al.\ 2000).  Has an unusually
``efficient'' PWN ($L_X \approx 0.25 \dot E$; Helfand et
al.\ 2003) and no radio pulse detection (Camilo et al., in preparation).

\subsection{Possible SNR associations with multibeam pulsars}

The most curious candidates are PSR~J1726--3530/G352.2--0.1 and
PSR~J1632--4818/G336.1--0.2 (Manchester et al.\ 2002).  The putative
SNRs are apparently coincident shells identified in MOST survey maps and
require further investigation.  Interestingly, both pulsars have $\tau_c
< 20$\,kyr and $P \approx 1$\,s --- two more examples of large-$B$,
long-$P$, youthful multibeam pulsars: 16 out of 43 with $\tau_c <
10^5$\,yr have $P>0.4$\,sec, compared to five out of 33 previously known!

\subsection{Possible EGRET associations with multibeam pulsars}

Two energetic multibeam pulsars that are coincident in projection with two
unidentified EGRET sources can plausibly power them (D'Amico et al.\ 2001).
A few other pairings are possibly real (e.g., Torres et al.\ 2001; Kramer
et al.\ 2003).  However, none of these cases are ironclad.  The difficulty
in assessing such identifications is illustrated by one small patch of sky
containing three energetic multibeam pulsars, two EGRET sources, one PWN,
one SNR, with substantial overlap --- and with no clear understanding of
the relationships, if any, between these objects (Roberts et al.\ 1999,
2001; Doherty et al.\ 2003).

\vspace{-2mm}
\section{Directed Searches}

\begin{table}[t]
\caption{Young pulsars discovered in directed searches.}
\label{tab:young}
\vspace{-3mm}
\begin{center}
\begin{tabular}{lccclc}\tableline
PSR      & $P$   & $\tau_c$  & $\dot E$ & Association & Refs \\
         & (ms)  & (kyr) & (ergs\,s$^{-1}$) &  &  \\
\tableline
J2229+6114 & 51 & 10 & $2\times10^{37}$ & 3EG J2229+6122 & (1) \\
J2021+3651 & 104 & 17 & $3\times10^{36}$ & 3EG J2021+3716 & (2) \\
J0205+6449 & 65 & 5.4 & $3\times10^{37}$ & SNR 3C~58 & (3) \\
J1124--5916 & 135 & 2.9 & $1\times10^{37}$ & SNR G292.0+1.8  & (4) \\
J1930+1852 & 136 & 2.9 & $1\times10^{37}$ & SNR G54.1+0.3 & (5) \\
J1747--2958 & 98 & 25 & $2\times10^{36}$ & ``Mouse'' PWN  & (6) \\
\tableline
\end{tabular}
\end{center}
\vspace{-2mm}
Refs: (1) Halpern et al.\ 2001; (2) Roberts et al.\ 2002; Hessels et al.\ 2004; (3) Murray et al.\ 2002; Camilo et al.\ 2002b; (4) Hughes et al.\ 2001; Camilo et al.\ 2002a; Hughes et al.\ 2003; (5) Lu et al.\ 2002; Camilo et al.\ 2002c; (6) Camilo et al.\ 2002d; Gaensler et al.\ 2004.
\end{table}

\vspace{-2mm}
\subsection{EGRET error boxes}

The ``method'' used in \S4.3 for evaluating the potential association of
serendipitously discovered energetic pulsars with positionally coincident
EGRET error boxes suffers from one fatal flaw: there is no guarantee
that the $\gamma$-ray source (which must produce substantial X-ray flux)
is not some as-yet-unidentified X-ray source within the error box,
corresponding to an as-yet-undetected pulsar.

The only way to be certain of an EGRET source--neutron star correspondence
(short of detection of $\gamma$-ray pulsations), is to survey the entire
EGRET box, methodically identifying all X-ray sources, until only one
remains otherwise unidentified that has characteristics typical of neutron
stars.  This method was used by Halpern et al. (2002) to identify the
magnetospherically active neutron star RX~J1836.2+5925 with 3EG~J1835+5918
--- even if, despite considerable efforts, no pulsations were detected.
The same method pursued by Halpern et al. (2001) led to the detection
of PSR~J2229+6114, the first object listed in Table~\ref{tab:young}.
With some differences, a broadly comparable method was used by Roberts
et al. (2002) to survey five EGRET GeV sources, leading to the detection
of PSR~J2021+3651.  In these cases, after extensive preliminary work,
pulsations were first detected in deep radio searches, and then also
in X-rays.  A further recent search along these lines was performed by
Becker et al. (2004) of a candidate neutron star within 3EG~J2020+4017
coincident with the $\gamma$-Cygni SNR.  The source turned out not to
be a neutron star, and deep radio searches for pulsations from the EGRET
source were also negative.  However, in this case only part of the EGRET
box was surveyed with sensitive X-ray observations, and it is possible
that an as-yet-undetected X-ray source/neutron star lurks within.

\vspace{-2mm}
\subsection{Pulsar Wind Nebulae}

Pulsations were detected from the third-through-fifth sources in
Table~\ref{tab:young} after they had been identified as clear PWNe
with central point sources via high-resolution X-ray observations.
Sensitive pulsation searches were then successful, in X-rays in the case
of PSR~J0205+6449 and at radio wavelengths for the other two pulsars.
Following the initial detections, all three now have pulsations detected
in both wavebands.  Another example of such a search is that for the
pulsar in the SNR CTA~1 (coincident with 3EG~J0010+7309).  In this
case, \textit{Chandra} imaging observations reveal the pulsar, but even
extremely constraining radio searches have failed to uncover pulsations
(Halpern et al.\ 2004).  The last source in the Table was thought to
be a PWN shaped by a bow shock caused by the supersonic motion of an
unseen pulsar through the interstellar medium.  A deep radio observation
confirmed this by detecting the pulsar J1747--2958.

In these examples the pulsars have very low radio luminosities ($L_{1400}
\equiv S_{1400} d^2 \sim 1$\,mJy\,kpc$^2$, with $S_{1400}$ the 1400\,MHz
flux density), suggesting that such limits must be attained in other
searches before alternative explanations for non-detection of pulsations
(e.g., beaming, radio-quietness) need be entertained.

\vspace{-2mm}
\section{ROSAT Isolated Neutron Stars}
\vspace{-1mm}

\begin{table}[t]
\caption{INSs and their periods.}
\vspace{-2mm}
\label{tab:ins}
\small
\begin{center}
\begin{tabular}{lcr}\tableline
Name (RX)     & $P$ (s)  & Notes/References\\
\tableline
J0420.0--5022 & 3.4 & $B \la 8\times 10^{13}$\,G; Haberl et al.\ 1999, 2004 \\
J0720.4--3125 & 8.4  & $B \la 10^{13}$\,G; Zane et al.\ 2002; Kaplan et al.\ 2002 \\
J0806.4--4132 & 11.4 & $B \la 3\times 10^{14}$\,G; Haberl \& Zavlin 2002 \\
J1308.6+2127  & 10.3 & Hambaryan et al.\ 2002; Haberl et al.\ 2003 \\
J1605.3+3249  & ?    & van Kerkwijk et al.\ 2004 \\
J1856.5--3754 & ?    & Ransom et al.\ 2002 \\
J2143.0+0654  & ?    & Zampieri et al.\ 2001 \\
\tableline
\end{tabular}
\end{center}
\vspace{-7mm}
\end{table}
\normalsize

These are bright (nearby) soft X-ray sources, with optical counterparts,
no SNR associations, and with no radio emission detected (e.g., Kaplan
et al.\ 2003; Johnston 2003).  Table~\ref{tab:ins} lists these sources
and shows that many now have a periodicity determined.  In a search for
other neutron stars in a \textit{ROSAT} survey, Rutledge et al.\ (2003)
estimate that there are $\la 67$ possible candidates.

\vspace{-2mm}
\section{Central Compact Objects}
\vspace{-1mm}

These are bright X-ray sources at the center of SNRs, thought to be
neutron stars.  They have broadly similar spectral, but different temporal
properties.  No radio emission has been detected from any of them (e.g.,
Gaensler et al.\ 2000).  No pulsations have been detected from the CCOs
in Cas~A, G266.2--1.2, RCW~103, or Pup~A.  Candidate CCOs have been
recently identified in G347.3--0.5 (Lazendic et al.\ 2003) and Kes~79
(Seward et al.\ 2003), and no radio pulsations have been detected from
them either (Camilo et al., in preparation).

The one exception to the non-detection of pulsations from CCOs is of
the neutron star in the SNR PKS~1209--51/52, with $P=424$\,ms (Zavlin et
al.\ 2000).  However, the spin behavior of this source is puzzling (e.g.,
Zavlin et al.\ 2004), and the last word on it has not yet been written
(see Pavlov et al.\ and De Luca et al., these proceedings).

\vspace{-2mm}
\section{Conclusions}
\vspace{-1mm}

Six of the nine youngest Galactic pulsars known (with $\tau_c <
10$\,kyr) were discovered in the past five years at both radio and X-ray
wavelengths.  Three more energetic pulsars were discovered in directed
searches of PWNe, with several more detected in the multibeam survey.
These advances point to the complementarity of undirected and directed,
as well as of radio and X-ray searches.

The median luminosity for the multibeam pulsars with $\tau_c < 10^5$\,yr
is $L_{1400} \approx 27$\,mJy\,kpc$^2$, compared to $L_{1400} \approx
60$\,mJy\,kpc$^2$ for those previously known.  However, an analysis of
deep radio searches of 23 PWNe (Camilo et al., in preparation) shows
that the bottom of the luminosity function for young pulsars is at least
one-to-two orders of magnitude below this value.

We now also recognize that some very young pulsars, like PSR~J1119--6127
(\S4.1; see also \S4.2), do not betray their presence via previously
detected PWNe.  They may lurk within otherwise seemingly empty shell
SNRs or where no SNR has yet been identified.  The birth rate of these
systems is not known.

The impact of these realizations toward pursuing a thorough understanding
of the population of young rotation-powered neutron stars is sobering:
a \textit{lot} more work remains to be done!  Much deeper undirected
searches should be done, and the very deepest possible searches of
promising targets must be accomplished.  New instrumentation (e.g.,
ALFA at Arecibo; the GBT) will be useful in this regard, and so will
the future \textit{GLAST} mission.  This work should be done, as much
as possible, while \textit{Chandra} and \textit{XMM} are still happily
collecting photons.

Taken as a whole, the many advances of the past five years in the areas of
AXPs, SGRs, CCOs, INSs, and ``standard'' pulsars help paint a fascinating
if still rather incomplete picture of the demographics and environments
of young neutron stars.  The search for pulsations remains a key aspect
of this discipline.

\acknowledgments 
I thank my many collaborators who have helped make the past three years
spent searching for young pulsars so fun and productive; the support of
NSF, NASA, and NRAO; and AJF, whose company and love have made the past
year unforgettable.


\end{document}